\documentclass[twocolumn,showpacs,preprintnumbers,amsmath,amssymb]{revtex4}

\usepackage{graphicx}
\usepackage{dcolumn}
\usepackage{bm}

\newcommand{\be}{\begin{equation}}
\newcommand{\en}{\end{equation}}
\newcommand{\bea}{\begin{eqnarray}}
\newcommand{\ena}{\end{eqnarray}}

\newcommand{\hbo}{\hbox to 1 true cm {\hfill } }

\begin{document}


\title{ $1/f$ noise from vortex-antivortex annihilation }

\author{Kurt Langfeld}

\affiliation{%
Insitut f\"ur Theoretische Physik, Universit\"at T\"ubingen \\
Auf der Morgenstelle 14, D-72076 T\"ubingen, Germany. \\
}%

\author{Dietmar Doenitz, Reinhold Kleiner, Dieter Koelle}

\affiliation{%
Physikalisches Insitut -- Experimentalphysik II, Universit\"at T\"ubingen \\
Auf der Morgenstelle 14, D-72076 T\"ubingen, Germany. \\
}%

\date{November 25, 2005}

\begin{abstract}
The magnetic flux noise induced by vortices in thin
superconducting films is studied. Rigid vortices in
thin films as well as pancake vortices in the pancake
gas regime are addressed. The vortex dynamics is
described by a Feynman path integral which fully
accounts for the balance between vortex entropy and
vortex energetics. We find that vortex pair creation
(annihilation) in the presence of pinning is a natural
source for noise with a $1/f$ spectrum.
\end{abstract}


\maketitle

Over the last decade it was noticed that the theory of strong interactions
(QCD) and type II superconductors share a common feature: it was
realized that the QCD vacuum sustains percolating
vortices~\cite{Greensite:2003bk}. The dynamics of the vortex texture
is thereby related to important properties of QCD such as
quark confinement. For instance, quark deconfinement at high temperatures
appears as vortex depercolation transition\cite{Engelhardt:1999fd}, and the
universality class of the transition can be anticipated by investigating the
vortex dynamics only~\cite{Langfeld:2003zi}.
It turned out~\cite{Engelhardt:1999wr} that a vortex model which
essentially incorporates the vortex entropy and an energy penalty for
vortex curvature already captures the dynamics of the confining
vortices. Below, we will apply this idea to describe vortices of
type II superconductors: In the case that the superconductor partition
function is dominated by the vortex entropy, the vortex dynamics is
described by a Feynman ``path integral'' which correctly accounts for the
vortex entropy. The only energy penalty is given for the vortex
mobility.

\bigskip
A central issue in the context of type II
superconductors is the magnetic flux noise and voltage
noise generated by the dynamics of vortices in
superconductors. This topic has been investigated both
theoretically and experimentally throughout the last
four decades, starting with the pioneering work by Van
Ooijen and Van
Gurp~\cite{VanOoijen65,VanGurp68,VanGurp69}. Early
work focused on so-called flux-flow noise
\cite{Clem81a} due to density and velocity
fluctuations of vortices driven by a bias current,
which causes voltage fluctuations in the flux flow
regime with a noise power spectrum often scaling with
frequency $f$ as $1/f^\alpha$ ($\alpha\approx 1 - 2$).
More recently, various groups have performed
experiments probing directly the magnetic flux noise
in superconductors which is sensed by nearby magnetic
field or flux
sensors\cite{Yeh84,Yeh91,Ferrari88,Rogers92,Placais94,Maeda02}.
Within such an approach, the magnetic flux noise at
low magnetic fields has also been detected by placing
a superconductor in close vicinity to a
superconducting quantum interference device (SQUID)
which senses directly the motion of vortices
\cite{Ferrari94}. Similarly, one can simply detect the
motion of vortices located in the SQUID structure
itself \cite{Clem05}, e.g., combined with imaging of
such vortices \cite{Doenitz04}. Such measurements in
low magnetic fields show also typically scaling of the
spectral density of flux noise as $1/f$.

\bigskip
Apart from the attempt to understand the origin and
mechanisms of such noise as a prerequisite for
improving the properties of superconducting devices,
the investigation of noise was also driven by the wish
to exploit vortex dynamics as a probe of the
superconducting state, in particular after the
discovery of the high transition temperature $T_c$
cuprate superconductors, which show very rich
phenomena associated both with the static and dynamic
properties of 'vortex
matter'~\cite{Bishop92,Blatter94,Brandt95,Crabtree97}.
An interesting aspect is related to the layered
structure of the cuprate superconductors and very thin
superconducting films, which presumably leads to
enhanced two-dimensional fluctuations near the
Kosterlitz-Thouless transition as determined from
simulations of $XY$-models with time-dependent
Ginzburg-Landau dynamics \cite{Houlrik94,Jonsson94}.

\bigskip
In this letter, the magnetic flux noise induced by a
vortex to a slot, similar to the situation of vortices
trapped in a SQUID washer \cite{Clem05}, is studied on
the basis of a statistical vortex model. The model
applies to the case of rigid vortices in thin films as
well as to pancake vortices \cite{Clem05a} in the
pancake gas regime close to $T_c$. Disregarding vortex
pair creation (annihilation), we find that the noise
spectrum shows a $1/\sqrt{f}$ behavior. Taking into
account vortex pair creation (annihilation), we
observe a $1/f$ law if the vortex pinning is active,
while the spectrum behaves like $1/f^2$ if pinning is
absent. This demonstrates that vortex pair creation
(annihilation) of pinned vortices can be a natural
source for $1/f$ noise.

\begin{figure*}
\includegraphics[height=7cm]{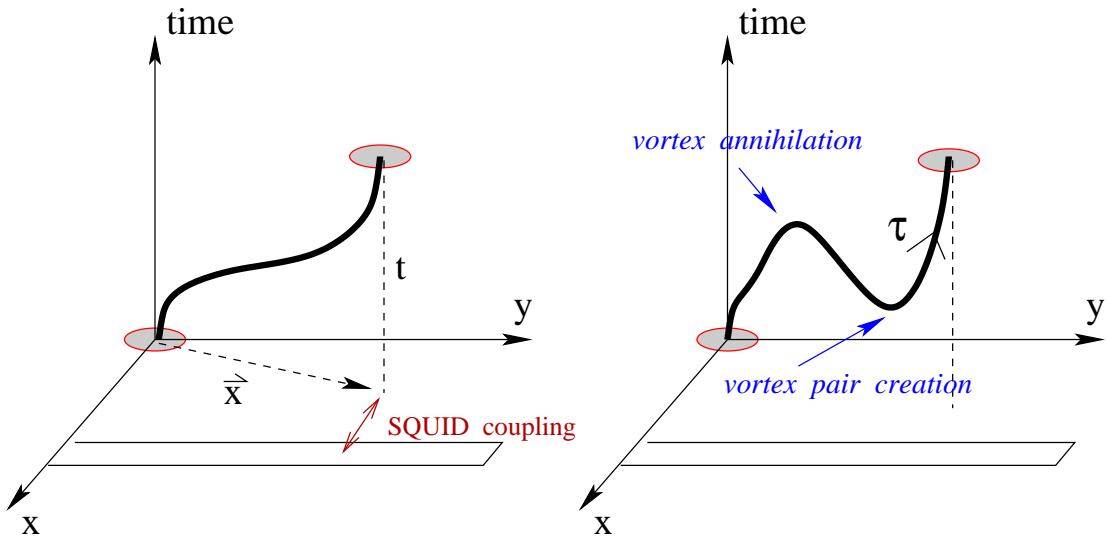}
\caption{\label{fig:1} Motion of single ``pancake''
vortex (left); Vortex pair creation and
vortex-antivortex annihilation (right). }
\end{figure*}

\bigskip
Let us consider a single ``pancake'' vortex which moves in
the 2-dimensional $xy$ plane. Vortex motion induces a
time dependent magnetic flux to the slot parallel to the $y$-axis
(see figure \ref{fig:1}).
The vortex ``zitterbewegung'' causes a frequency  dependent noise
to the SQUID, which measures the flux induced to the slot.
The derivation of the noise spectrum $S(f)$ will be our goal
below. The motion of the vortex is described in the present case by a
2-dimension vector $\vec{x}(t)$ which depends on the time $t$.
In order to describe the pinning of the vortex to the origin $\vec{x}=0$,
we constrain the vortex motion by the {\tt weak pinning condition}:
\be
\frac{1}{t} \int _0^t dt^\prime \; \vec{x}(t^\prime) \; = \; 0.
\label{eq:1}
\en
The latter condition implies that the ``center of mass'' of the
vortex is tied to the origin of the coordinate system. The
quantity of primary interest is the probability distribution
\be
P(\vec{x}_0,\vec{x};0,t) \; = \; \langle \vec{x}(0) \vec{x}(t) \rangle
\label{eq:2}
\en
of finding a vortex at $ \vec{x} $ at time $t$, when the vortex
was located at  $ \vec{x}_0 $ at time $t=0$. The average in
(\ref{eq:2}) is taken over all vortex world lines which satisfy
the pinning condition (\ref{eq:1}) and the boundary conditions
$(\vec{x}_0,0)$ and $(\vec{x},t)$.
The noise spectrum is obtained from the probability distribution
by $(\omega = 2 \pi f)$
\be
S(\omega ) \; = \; \int dt \; d^2x_0 \; d^2x \; T_d (\vec{x}) \;
P(\vec{x}_0,\vec{x};0,t) \; e^{i \omega t } \; .
\label{eq:3}
\en
Thereby, $T_d(\vec{x})$ is the transfer function which describes
the coupling of the magnetic field of the vortex located at
$\vec{x}=(x,y)$ at time $t$ to the slot, which is loacted at distance
$d$ from the $y$-axis.
A simple model for the transfer function is provided by the choice
\be
T_d(\vec{x}) \; = \; \delta (x-d) \; , \hbo \vec{x} = (x,y) \; .
\label{eq:3a}
\en
Thereby, it is assumed that only ``pancake'' vortices which
possess overlap with the slot induce a signal.
The $\vec{x}$ integration in (\ref{eq:3}) sums up
all contrubtions to the slot, whereas the $\vec{x}_0$ integration can
be understood as an average over all initial conditions of the
vortex.

\bigskip
In the following, we will assume that the partition function of the
vortex matter is solely given by the interplay between vortex
entropy and vortex energy. In this regime, the dynamic of the vortices
is captured by a statistical model. Correlation functions
are obtained by averaging over ``pancake'' vortex ensembles:
The probability distribution is effectively described in terms of the
``Feynman'' path integral
\bea
P(\vec{x}_0,\vec{x};0,t) &= &  {\cal N}
\int {\cal D} \vec{x} (t) \; \delta \biggl( \frac{1}{t}
\int _0^t dt^\prime \;  \vec{x}(t^\prime)  \biggr)
\label{eq:4} \\
&& \exp \left\{
- \int _0^t dt^\prime \; \left[ \frac{m}{2} [\dot{ \vec{x} }
(t^\prime)]^2 \right] \right\} \; ,
\nonumber
\ena
where $m$ is the only parameter of the model. $m$ parametrizes the
mobility of the ``pancake'' vortex moving in the 2d plane.
All vortex worldlines, which contribute to the path integral
(\ref{eq:4}) start at $\vec{x}_0$ at time $t=0$ and end at
$\vec{x}$ at time $t$. The normalization constant $ {\cal N} $ can
be obtained from the condition
$
\int d^2x \; P(\vec{x}_0,\vec{x};0,t) \; = \; 1 \; .
$
Let us calculate the auxilliary quantity (the dependence on
$\vec{x}_0$, $\vec{x}$ is suppressed)
\bea
Z(\vec{\alpha}, t) &=& \int {\cal D} \vec{x} (t) \; \exp \left\{ i
\frac{\vec{\alpha} }{t}
\int _0^t dt^\prime \;  \vec{x}\left(t^\prime \right)  \right\}
\label{eq:5} \\
&& \exp \left\{
- \int _0^t dt^\prime \; \left[ \frac{m}{2} [\dot{ \vec{x} }
(t^\prime)]^2 \right] \right\} \; ,
\nonumber
\ena
$Z(0,t)$ corresponds to the probability distribution of freely moving
vortices, while integration over $\vec{\alpha} $ installs the weak
pinning condition (\ref{eq:1}), i.e.,
\be
P(\vec{x}_0,\vec{x};0,t) \; = \; \int \frac{d^2\alpha }{(2\pi )^2} \;
Z(\vec{\alpha},t) \; .
\label{eq:6}
\en
It is tedious, but straightforward to calculate the function
$Z(\vec{\alpha}, t)$ in (\ref{eq:5}). For this purpose, the
time interval $[0,t]$ is divided into $N$ pieces of equal distance:
\bea
t^\prime _n&=&  \frac{t}{N}\; n \; \hbo n=0 \ldots N \; ,
\nonumber \\
\vec{x}(t^\prime_n ) &=:& \vec{x}_n \; ,  \hbo \vec{x}_N \equiv  \vec{x} \; .
\nonumber
\ena
The discretized version of the path integral (\ref{eq:5})
is given by
\bea
Z(\vec{\alpha}, t) &\rightarrow & t^{-N/2}
\int d\vec{x}_1 \ldots d\vec{x}_{N-1}  \; \exp \biggl\{ i
\frac{\vec{\alpha} }{N} \sum _{n=0}^N \vec{x}_n
\nonumber \\
&-& \frac{m \, N}{2t} \sum _{n=0}^{N-1} (\vec{x}_{n+1} -
\vec{x}_n )^2 \biggr\} \; ,
\nonumber
\ena
where the prefactor of the functional integral arises from the measure
${\cal D}x(t)$.
After performing the Gaussian integrals and taking the limit
$N \rightarrow \infty$, we obtain up to an unimportant
numerical factor
\bea
Z(\vec{\alpha}, t) &\propto &
\exp \left\{ - \frac{m}{2t} \left( \vec{x} -  \vec{x}_0 \right)^2
\right\}
\label{eq:7} \\
&& \exp \left\{ - \frac{t}{24m} \vec{\alpha }^2
+ \frac{i}{2}  \vec{\alpha } (  \vec{x} +  \vec{x}_0 ) \right\} \; .
\nonumber
\ena
Setting $\vec{\alpha }=0$ in the last expression, we find that
$Z$ only depends on $ \vec{x} -  \vec{x}_0$ reflecting translation
invariance in the absence of the pinning center at the origin.
Extending the correlation function to negative times, we replace
$t$ in (\ref{eq:7}) by $\vert t \vert $. Inserting (\ref{eq:7})
in (\ref{eq:3}) and introducing the relative
momentum $\vec{q}$
which is the conjugate variable to $ \vec{x} -  \vec{x}_0 $,
the integration over $t$, $\vec{x}_0$ and $\alpha $ can be performed
yielding
\be
S(\omega ) \; \propto  \; \int \frac{d^2q}{(2\pi )^2} \; d^2x \;
\Re \left\{ \frac{1}{ \frac{\vec{q}^2}{6m} + i \omega } \right\} \;
T_d (\vec{x}) \; e^{i \vec{q} \vec{x} } \; .
\label{eq:8}
\en
For the idealized transfer function in (\ref{eq:3a}), we finally find
for the case $\sqrt{6m \omega } \, d \ll 1$ that
\be
S(\omega ) \; \propto  \; \sqrt{ \frac{6m}{\omega } } \; .
\label{eq:9}
\en
Hence, statistical ensembles, consisting out of single and weakly pinned
vortices, cannot account for the $1/f$ type spectrum.

\bigskip
In order to allow for the merge of a vortex antivortex or for
the vortex pair creation in the vortex statistical ensemble,
the vortex world line is addressed by an implict parameterization:
$$
 \vec{x}(\tau ) = (x(\tau ), y(\tau)) \; , \; \; \; \;
t(\tau) \; , \; \; \; \;
\underline{x} := (x,y,t) \; ,
$$
where the parameter $\tau $ will be called ``proper time'' below.
If $t(\tau )$ is a non-monotonic function, vortex pair creation and
annihilation is take into account. The weak pinning condition (\ref{eq:1})
is generalized to an average over the vortex world line
\be
\frac{1}{\tau } \int _0^\tau d\tau ^\prime \; \vec{x}(\tau^\prime) \; = \; 0.
\label{eq:10}
\en
The vortex statistical ensemble is now approached by means
of the path integral
\bea
Z_{CA}(\alpha, t) &=& \int {\cal D} \underline{x} (\tau ) \; \exp \left\{ i
\frac{\vec{\alpha} }{\tau }
\int _0^\tau  d\tau^\prime \;  \vec{x}\left(\tau^\prime \right)  \right\}
\label{eq:11} \\
&& \exp \left\{
- \int _0^\tau d\tau^\prime \; \left[ \frac{m}{2} \dot{ \vec{x} } ^2
+ \frac{\mu}{2} \dot{ t }^2 \right] \right\} \; ,
\nonumber
\ena
where the dot denotes the derivative with $\tau $. If $\mu $ is
small, $t(\tau )$ possesses many non-monotonic parts. This implies
that $1/\mu $ is related to the vortex pair creation rate. The
ensemble average must be taken over all vortex world lines starting at
$\underline{x}_0 := (x_0,y_0,0) $ and ending at $\underline{x} := (x,y,t)$.
There is no restriction for the proper time $\tau $ implying that we will
take the limit $\tau \rightarrow \infty$ below. A calculation,
analogous to that in the previous subsection, yields
\bea
Z_{CA}(\vec{\alpha}, t) &\propto & \int _0^\infty d\tau \;
\exp \left\{ - \frac{m}{2\tau} \left( \vec{x} -  \vec{x}_0 \right)^2
- \frac{\mu }{2\tau} t^2 \right\}
\nonumber  \\
&& \exp \left\{ - \frac{\tau}{24m} \vec{\alpha }^2
+ \frac{i}{2}  \vec{\alpha } (  \vec{x} +  \vec{x}_0 ) \right\} \; .
\label{eq:12}
\ena
After switching to the momentum space, the $\tau $ integration can be
performed, i.e.,
\be
Z_{CA} (\vec{\alpha}, t)\propto \int \frac{d^3q}{(2\pi)^3} \;
\frac{ e^{i \vec{q}(\vec{x}-\vec{x_0})} e^{i q_0 t}
e^{i/2 \, \vec{\alpha}(\vec{x}+\vec{x_0})} }{
\frac{\vec{q}^2 }{2m} + \frac{q_0^2 }{2\mu} +
\frac{\vec{\alpha }^2 }{24m} } \; .
\label{eq:13}
\en
The noise spectrum $S(\omega )$ is now easily obtained by
inserting (\ref{eq:13}) in (\ref{eq:3}). Setting $\vec{x}=(x,y)$
and  $\vec{q}=(q_1,q_2)$, the $\vec{x}_0$, $\vec{\alpha }$,
$q_0$ and $t$ integration is straightforward:
\be
S(\omega ) \propto \int \frac{d^2q}{(2\pi)^2} \; d^2x \; T_d(x) \;
\frac{ e^{i 2 \vec{q} \vec{x} } }{
\frac{ 2 \vec{q}^2 }{3m} + \frac{\omega ^2 }{2\mu} } \; .
\label{eq:14}
\en
For the idealized transfer function in (\ref{eq:3a}), our final result
is
\be
S(\omega ) \propto \; \exp \left\{ - \sqrt{\frac{3m}{\mu } }
\omega \, d \right\} \; \frac{ \sqrt{ 3m \mu } }{ 2\mu \, \omega } \; .
\label{eq:15}
\en
This is the central result of our letter: allowing for vortex
pair creation and annihilation, the noise spectrum shows
$1/\omega $ dependence at small frequencies if the vortex is pinned
sufficiently close ($\omega d \ll 1$) to the slot.

\bigskip
How does pinning affect the noise spectrum? Let us
consider the vortex statistical ensembles which
include vortex pair creation and annihilation, but
where the pinning constraint is absent. The noise
spectrum is given in this case by \be S(\omega ) \;
\propto \; \int dt \; d^2x_0 \; d^2x \; T_d (\vec{x})
\; Z_{CA}(0,t) \; e^{i \omega t } \; . \label{eq:16}
\en Note that we have set $\vec{\alpha }=0$ in the
auxiliary function $Z_{CA}(\vec{\alpha} ,t)$ rather
than integrating over $\vec{\alpha }$. Using
(\ref{eq:13}), the frequency dependent part of
$S(\omega )$ is given by $ S(\omega ) \propto 1/\omega
^2 $. Obviously, the pinning of the ``pancake''
vortices is important to obtain a $1/f$ noise
spectrum.

We thank T.~Tok for valuable comments on the
manuscript. This work was supported by a grant from
the Ministry of Science, Research and the Arts of
Baden-W\"urttemberg (Az: 24-7532.23-19-18/1).

\bibliography{kurtref}

\end{document}